\documentclass[aps,eqsecnum,groupedaddress,a4paper,floatfix,twocolumn]{revtex4}
\usepackage[T1]{fontenc}
\usepackage{mathptmx}
\usepackage{amsfonts}
\usepackage{amssymb}
\usepackage{mathrsfs}
\usepackage{graphicx}
\usepackage{fancyhdr}

\newcommand{\XDOI}[1]{\href{http://dx.doi.org/#1}{doi:#1}}
\newcommand{\XARXIV}[1]{\href{http://arxiv.org/abs/#1}{arXiv:#1}}

\begin{document}
\title{
The Hall Effect and ionized impurity scattering in Si$_{(1-x)}$Ge$_x$
}

\author{P. Kinsler\cite{e-mail}} 
\affiliation{
  Department of Physics, Imperial College,
  Prince Consort Road,
  London SW7 2BW, 
  United Kingdom.
}
\author{W.Th. Wenckebach}
\affiliation{
Department of Applied Physics, Technical University Delft, 
Lorentzweg 1, 2628 CJ DELFT, The Netherlands. 
}
\date{\today}

\lhead{
\includegraphics[height=5mm,angle=0]{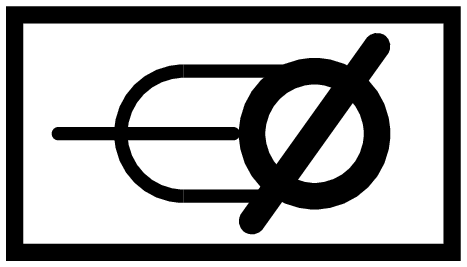}~~SIGELIIM}
\chead{~}
\rhead{Dr.Paul.Kinsler@physics.org\\
http://www.kinsler.org/physics/}

\begin{abstract}

Using Monte Carlo simulations, we demonstrate that including 
ionized impurity scattering in models of Si$_{(1-x)}$Ge$_x$
is vital in order to predict the correct Hall parameters.
Our results show good agreement with the experimental
data of Joelsson et.al. 
\cite{Joelsson-FNH-1997jap}.

\end{abstract}

\maketitle
\thispagestyle{fancy}


\noindent
\emph{Published as: \\
J. Appl. Phys. 94, 7159 (2003), \\
\XDOI{10.1063/1.1622994}.
}

\section{Introduction}

There is much current interest in using Si/Ge
strained layer superlattices and multiple quantum wells to fabricate
optoelectronic and electronic devices --  one example being better high
frequency performance for heterojunction bipolar transistors. 
In order to
design effective devices we need to know the carrier concentration $N_H$ and
the conduction mobility $\mu_c$, which are usually obtained using Hall
measurements.  However, to extract $N_H$ and $\mu_c$ from the experimental
results, we need to know the Hall coefficient $r_H$ -- but experiments in
Si$_{(1-x)}$Ge$_x$ \cite{Chen-LBSH-1994apl,Joelsson-FNH-1997jap} have shown
$r_H$ has values that cannot be accounted for by existing attempts at
modelling (see e.g.  \cite{Dijkstra-1997tudelft,Dijkstra-W-1999jap}).  Also,
the Hall factor $r_H$ allows an interesting test of the accuracy of the band
parameters because it is quite sensitive to both the non-parabolicity and
anisotropy of the bands.  

The reason that Hall data from existing experiments
\cite{Joelsson-FNH-1997jap,Chen-LBSH-1994apl} is not well explained by
existing models is most probably either because (a) these models used a
linearly interpolated bandstructure, or (b) they omitted the important effects
of ionised impurity scattering.  Alternatives to a linearly interpolated
bandstructure have been given by Rieger and Vogl \cite{Rieger-V-1993prb},
using an empirical psuedopotential approach, and Hayes suggests a similar
method based on results of a parallelized Car-Parrinello scheme
\cite{Haye-1998tudelft}.  Some other papers discussing models and
bandstructure parameters for SiGe alloys are
\cite{Bufler-GMFK-1998jvst,Fischetti-L-1996jap,Liou-WC-1994jap}.

Here we attack the second of these shortcomings, by calculating the Hall
parameters using a comprehensive Monte Carlo simulation of hole motion in
Si$_{(1-x)}$Ge$_x$ strained to a silicon
substrate.  These simulations are aided by the incorporation of a new
optimised scheme for treating ionised impurity
scattering\cite{Wenckebach-K-2002cpc}.  Our results show that the inclusion of
ionised impurity scattering sucessfully explains the experimental data,
leaving the bandstructure interpolation as a secondary effect.

The paper is organised as follows: section \ref{SMC} describes the Monte Carlo
model used in the mobility and Hall calculations; section \ref{Sresults}
compares the results of the simulations for linearly interpolated band
parameters with experiments.  Finally in section
\ref{Sconclusions} we present our conclusions.


\section{The Monte Carlo Model}\label{SMC}

Our Monte Carlo simulations\cite{MonteCarlo} of this system include a full
$k.p$ band structure calculation \cite{Kane-1966ss} and all important
scattering processes: optical and accoustical deformation potential, alloy,
and (in particular) ionised impurity
scattering\cite{Ridley-QPS,Reggiani-1985hts}.  We treat the effect of the
electric and magnetic fields classically, giving the holes continuous
trajectories in $k$-space -- comparison with experiment for p-Ge systems
indicates that this approximate treatment is adequate for the field strengths
we consider.  We used a similar Monte Carlo model recently for studies of hot
hole lasers\cite{Kinsler-W-2001jap,Kinsler-W-2000icps}.  All simulations are
done at a lattice temperature of 300K.  They follow the progress of a single
hole through a large number of scatterings (typically $\sim 10^6 \times
100$), with the ergodic theorem being used to justify the use of the
time-average as an ensemble average.  Often a very large number of scatterings
was needed in order to make sure the statistical errors in the Hall parameter
were small enough.  Our material parameters are listed in table
\ref{Table-SiGe}.

The Monte Carlo simulations used the standard overestimation technique where
for each scattering process the post-scattering direction of the hole was
chosen at random, and the differential scattering rate was overestimated by an
isotropic rate just higher than its maximum value.  Normally
this would be very slow for the ionised impurity scattering, especially at low
impurity concentrations.  Consequently, we use an optimised scheme
\cite{Wenckebach-K-2002cpc} that weights the choice of scattering angle by the
angular dependence of the scattering rate, and thus avoid generating a large
proportion of inefficient overestimations.


\section{Results}\label{Sresults}

The Hall co-efficient $r_H$ is very sensitive to both the shape and
non-parabolicity of the bandstructure, as well as the anisotropy of the
scattering processes.  This means that the role of ionised impurity scattering
can be crucial --  not only the presence or absence of impurities makes a
difference; different impurity concentrations can produce Hall factors that
vary with alloy concentration in both quantatively and qualitatively different
ways.  As a result we need to compare our simulation results with experiment
to ascertain whether our model is sufficiently accurate and include
all the necessary important features.

Experimental results for the Hall factor in relaxed SiGe have been reported by
Chen et.al. \cite{Chen-LBSH-1994apl} for impurity concentrations of about $n
\sim 10^{17}$cm$^{-3}$; and in SiGe strained to a Si substrate by Joelsson
et.al. \cite{Joelsson-FNH-1997jap} for $n \sim 10^{18}$cm$^{-3}$.  A
comparison of these experimental results with Monte Carlo simulations has been
done by Dijkstra
\cite{Dijkstra-1997tudelft,Dijkstra-W-1999jap}, whose 
simulations used linearly interpolated band parameters and neglected ionised
impurity scattering. We have done equivalent simulations to those of Dijkstra,
and include the results here.

To calculate the Hall factor, we use the efficient estimator of the Hall 
mobility $\mu_H$ presented by \cite{Dijkstra-1997tudelft,Dijkstra-W-1997apl}, 
\begin{eqnarray}
\mu_H
&=&
\frac
{
  \left< \vec{v}
         \cdot 
         \left( \vec{r} \times \vec{B} \right) 
  \right>
}
{
  \left< \left( \vec{v} \times \vec{B} \right) 
         \cdot 
         \left( \vec{r} \times \vec{B} \right)
  \right>
}
,
\end{eqnarray}

and then since $r_H = \mu_H / \mu_c$, the drift mobility is
\begin{eqnarray}
\mu_c
&=&
\frac{e}{k_B T}
\frac{1}{2\left|\vec{B}\right|^2}
  \left< \left( \vec{v} \times \vec{B} \right) 
         \cdot 
         \left( \vec{r} \times \vec{B} \right)
  \right>
.
\end{eqnarray}

We did our simulations with no electric field, and a magnetic field in the
$[001]$ direction, in accordance with experiment; magnetic field strengths of
$B= 0.30$, $1.00$, or $2.00$T were investigated.  We used a 
strain equivalent to that as if the
Si$_{1-x}$Ge$_x$ alloy had been grown as an epitaxial layer on a (001)Si 
substrate.  This means the usual strain parameter $\epsilon_{zz} > 0$, since
the lattice parameter of Si is smaller than that for Ge.  

Our  alloy scattering potential was $U_0 = 0.51$eV.  
Note that some authors
\cite{Fischetti-L-1996jap,Liou-WC-1994jap}, on the basis of fits to mobilities
and other transport parameters, prefer a stronger alloy scattering potential
($\sim 1$eV); other discussions of the effect of alloy potential (and high
impurity concentrations) can be found in
\cite{Briggs-WH-1998sst,Briggs-WH-1998sst2}.  We do not expect our results for
$r_H$ to be particularly sensitive to the strength of the alloy potential, as
alloy scattering is isotropic.  This assumption was confirmed by a set of
simulations on undoped SiGe using the Monte Carlo program of
Dijkstra\cite{Dijkstra-1997tudelft}.  However, we would expect the drift
mobility to diminish for higher concentrations of germanium is a stronger
alloy scattering potential were chosen.

\begin{figure}
\includegraphics[angle=-90,width=0.90\columnwidth]{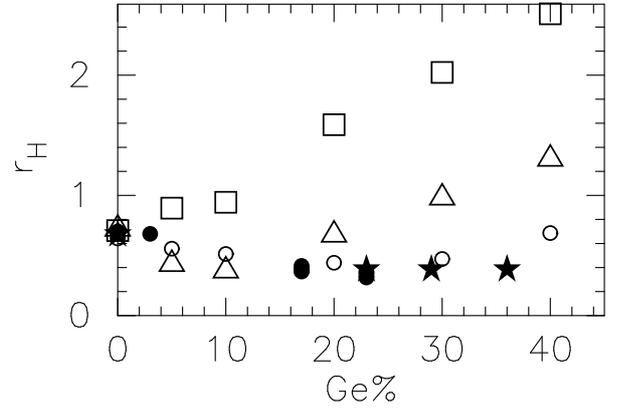}
\caption{ 
\label{Graph-compexpB0030}
Hall coefficient comparison with experiment\cite{Joelsson-FNH-1997jap}
for strained Si$_{1-x}$Ge$_x$ alloy, 
at $B=0.30$T, as a function of  Ge fraction. 
{\bf Monte Carlo:} 
$\Box = ${\em undoped}; 
$\circ = 3.16 \times 10^{18}$cm$^{-3}$ ; 
$\triangle =$ {\em undoped 2T}; 
{\bf Experiment\cite{Joelsson-FNH-1997jap}:} 
$\bullet = 2.08 $--$ 2.98 \times 10^{18}$cm$^{-3}$; 
$\bigstar = 6.59 $--$ 7.65 \times 10^{18}$cm$^{-3}$; 
Error bars are not included because the symbols become obscured, but
all simulation errors are listed in \protect{table \ref{Table-HallC0}}.
}
\end{figure}

Comparison with  experimental results can be seen on fig. 
\ref{Graph-compexpB0030}, where those for $\sim 2.5 \times 10^{18}$cm$^{-3}$
($\bullet$) and $\sim 7  \times 10^{18}$cm$^{-3}$ ($\bigstar$) show a Hall
coefficient decreasing as the Ge content increases.  These clearly do not
agree with our undoped Monte Carlo simulations at 0.3T ($\square$), although
Monte Carlo simulations at the higher field of 2T ($\triangle$) do also have
dip at low-Ge fraction.  This dip at 2T has been previously reported for Monte
Carlo simulations without ionised impurity scattering
\cite{Dijkstra-1997tudelft,Dijkstra-W-1999jap}.  However, the 2T simulations
have a minimum $r_H$ at 15\% Ge, with a subsequent increase --
at 40\% Ge $r_H$ is well above 1.  This is contrary to experiments,
 where $r_H$ continues to decrease up to 36\% Ge.

\begin{figure}
\includegraphics[angle=-90,width=0.90\columnwidth]{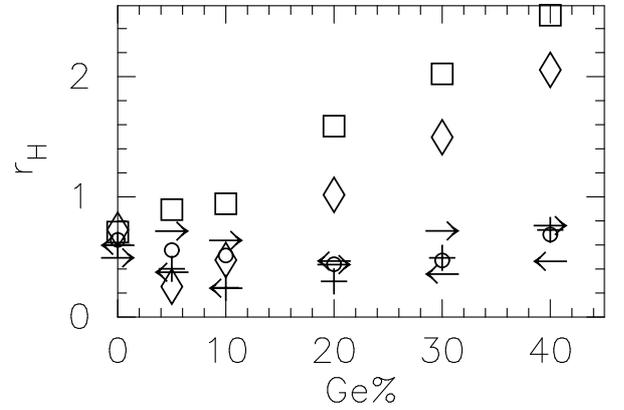}
\caption{ 
\label{Graph-compareB0030}
Hall coefficients for strained 
Si$_{1-x}$Ge$_x$ alloy, 
at $B=0.30$T, as a function of  Ge fraction. 
{\bf Monte Carlo:} 
$\Box = ${\em undoped}; 
$\Diamond = 10^{15}$cm$^{-3}$; 
$+ = 10^{17}$cm$^{-3}$; 
$\leftarrow = 10^{18}$cm$^{-3}$;
$\circ = 3.16 \times 10^{18}$cm$^{-3}$;
$\rightarrow =  10^{19}$cm$^{-3}$.
Error bars are not included because the symbols become obscured, but
are listed in  \protect{table \ref{Table-HallC0}}.
}
\end{figure}

However, it is clear that for undoped SiGe (fig. \ref{Graph-compexpB0030},
$\Box$), there is no dip like that seen in experiments on doped material --
hence, results for undoped SiGe do {\em not} explain the experiment.  Our
simulations with $B=0.3$T and $N_a = 3.16 \times
10^{18}$cm$^{-3} (\circ$) are shown on fig. \ref{Graph-compexpB0030}
(also repeated on fig. \ref{Graph-compareB0030}). 
These can be seen to match remarkably well with the experimental data taken at
$N_a = 2.5$ and $7\times 10^{18}$cm$^{-3}$, making it clear 
that it is important to include the ionised impurity
scattering in a model intended to reproduce the experimental value of $r_H$.

\begin{figure}
\includegraphics[angle=-90,width=0.90\columnwidth]{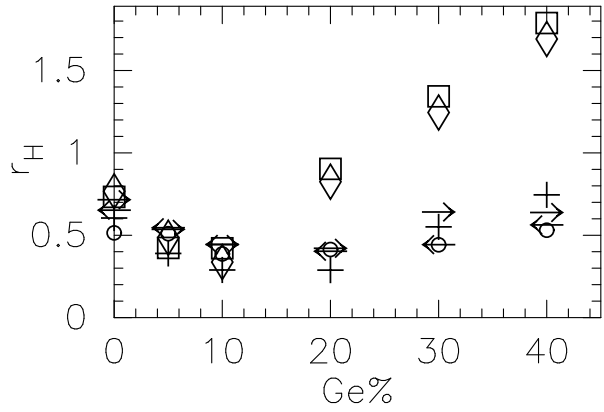}
\caption{ 
\label{Graph-compareB0100}
Hall coefficients for strained Si$_{1-x}$Ge$_x$ alloy, 
at $B=1.00$T, as a function of Ge fraction. 
{\bf Monte Carlo:} 
$\Box = ${\em undoped}; 
$\Diamond = 10^{15}$cm$^{-3}$; 
$+ = 10^{17}$cm$^{-3}$; 
$\leftarrow = 10^{18}$cm$^{-3}$;
$\circ = 3.16 \times 10^{18}$cm$^{-3}$;
$\rightarrow =  10^{19}$cm$^{-3}$.
Error bars are not included because the symbols become obscured, but
are listed in \protect{table \ref{Table-HallC0}}.  They are 
smaller for lower doping concentrations, are typically
smaller than $\pm0.1$, and are no greater than $\pm0.211$. 
}
\end{figure}

\begin{figure}
\includegraphics[angle=-90,width=0.90\columnwidth]{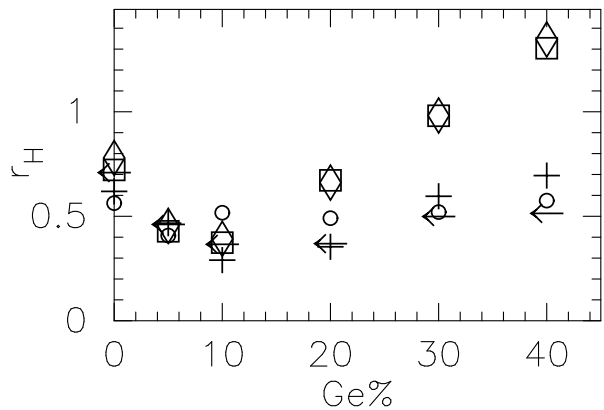}
\caption{ 
\label{Graph-compareB0200}
Hall coefficients for strained Si$_{1-x}$Ge$_x$ alloy, 
at $B=2.00$T, as a function of Ge fraction. 
{\bf Monte Carlo:} 
$\Box = ${\em undoped}; 
$\Diamond = 10^{15}$cm$^{-3}$; 
$+ = 10^{17}$cm$^{-3}$; 
$\leftarrow = 10^{18}$cm$^{-3}$;
$\circ = 3.16 \times 10^{18}$cm$^{-3}$;
$\rightarrow =  10^{19}$cm$^{-3}$.
Error bars are not included because the symbols become obscured, but
are listed in  \protect{table \ref{Table-HallC0}}.  They are 
smaller for lower doping concentrations, are typically
smaller than $\pm0.1$, and are no greater than $\pm0.192$.
}
\end{figure}

We can conclude that when impurity scattering dominates, $r_H$ stays low up to
high Ge concentrations, a case which contrasts with that studied by Dykstra,
who did not include impurity scattering.  Our results thus agree much better
with experiment than his -- indeed, the importance of impurity scattering was
suggested by Dykstra as a reason for the discrepancy he noticed.
The full range of $r_H$ values from the 
Monte Carlo results are shown on figs.
\ref{Graph-compareB0030}, \ref{Graph-compareB0100}, \ref{Graph-compareB0200}
for a range of $B$ values.  These emphasise the role of impurity scattering 
-- the $r_H$ values in 
undoped and low doping
simulations are significantly higher than those for higher doping.


\section{Conclusions}\label{Sconclusions}

Comparison of our Monte Carlo simulation results for doped strained
Si$_{(1-x)}$Ge$_x$ with experimental results \cite{Joelsson-FNH-1997jap} have
clearly shown that ionised impurity scattering has an important role in
determining the Hall parameter $r_H$.  This was shown by simulations which
without ionised impurity scattering failed to give agreement with experiment,
but with it gave good agreement.  This was despite using a linearly
interpolated set of band parameters, so clearly the interpolation scheme is
less important than the impurity scattering mechanism.

Following on from this work, it would be good to reduce the statistical
errors in the Hall co-efficients further.  This would enable us to
make comparisons between different interpolation schemes for the 
Si$_{(1-x)}$Ge$_x$ parameters.  Also, the behaviour of the Hall
parameter in relaxed Si$_{(1-x)}$Ge$_x$ would be interesting to
consider.




\clearpage 

\begin{table}
\begin{center}
\begin{tabular}{|r l|c|c|c|}
\hline
Parameter & & Si & Ge & Units \\
\hline
\hline
valence-band structure &   L     & -5.53     &    -30.53  &     dimensionless  \\  
                       &   M     & -3.64     &     -4.64  &     dimensionless \\
                       &   N     & -8.75     &    -33.64  &     dimensionless \\
\hline
 spin-orbit splitting  &  $D_{lo}$   & 0.044     &0.300  &     eV \\
 deformation potential &    a    & 2.1       &2.0    &     eV \\
                       &   b     & -2.2      &-2.1   &     eV \\
                       &   d     & -5.3      &-7.0   &     eV \\
                       &  $d_0$    & 29.3      &40.0   &     eV \\
\hline
 lattice constant      &  $a_0$    &  5.43095 & 5.6579 &    Angstrom \\
 elastic stiffness     &  $c_{11}$ & 16.56    &13.064  & $10^{11}$ dyn/cm$^2$\\
                       &  $c_{12}$ &  6.39    & 4.885  & $10^{11}$ dyn/cm$^2$\\
                       &  $c_{44}$ &  7.95    & 6.857  & $10^{11}$ dyn/cm$^2$\\
\hline
 mass density          &  $\rho$      &  2.328   &  5.3243     &   g/cm$^3$ \\
 optical phonon energy &  $E_{op}$    &  0.063   &  0.037      &   eV \\
 dielectric constant   &  $\epsilon$  & 11.90    & 15.90       &    \\
\hline
\end{tabular}
\end{center}
\caption{\label{Table-SiGe} Material parameters for silicon and germanium,
as used in our simulations. They are similar to those used by 
Hinckley and Singh\cite{Hinkley-S-1990}.}
\end{table}

\newpage

\begin{table}
 \begin{center}
\begin{tabular}{|r|r||rcr|rcr|rcr|rcr|rcr|rcr|}
\hline
\multicolumn{20}{|c|}{  Hall co-efficient $r_H$            } \\
\hline
 B (T)             &  Ge\%                                     & \multicolumn{18}{|c|}{$\log_{10} N_A$ (cm$^{-3}$)} \\
\hline 
          &            &       0.0 &&&       15.0 &&&       17.0 &&&       18.0 &&&       18.5 &&&       19.0 &&               \\
 \hline
   0.30 ~&    40 ~  &       2.417&$\pm$&     0.032  &       2.056&$\pm$&     0.052  &       0.725&$\pm$&     0.071  &       0.464&$\pm$&     0.118  &       0.687&$\pm$&     0.092  &       0.761&$\pm$&     0.161  \\
   0.30 ~&    30 ~  &       1.820&$\pm$&     0.043  &       1.496&$\pm$&     0.075  &       0.493&$\pm$&     0.091  &       0.358&$\pm$&     0.132  &       0.471&$\pm$&     0.115  &       0.717&$\pm$&     0.202  \\
   0.30 ~&    20 ~  &       1.157&$\pm$&     0.039  &       1.017&$\pm$&     0.101  &       0.297&$\pm$&     0.124  &       0.466&$\pm$&     0.151  &       0.440&$\pm$&     0.133  &       0.494&$\pm$&     0.196  \\
   0.30 ~&    10 ~  &       0.456&$\pm$&     0.022  &       0.475&$\pm$&     0.139  &       0.244&$\pm$&     0.164  &       0.240&$\pm$&     0.189  &       0.513&$\pm$&     0.164  &       0.637&$\pm$&     0.314  \\
   0.30 ~&     5 ~  &       0.408&$\pm$&     0.013  &       0.253&$\pm$&     0.160  &       0.402&$\pm$&     0.184  &       0.373&$\pm$&     0.211  &       0.557&$\pm$&     0.184  &       0.715&$\pm$&     0.349  \\
   0.30 ~&     0 ~  &       0.737&$\pm$&     0.011  &       0.726&$\pm$&     0.210  &       0.601&$\pm$&     0.196  &       0.597&$\pm$&     0.226  &       0.642&$\pm$&     0.200  &       0.493&$\pm$&     0.313  \\
 \hline
   1.00 ~&    40 ~  &       1.788&$\pm$&     0.016  &       1.691&$\pm$&     0.029  &       0.745&$\pm$&     0.035  &       0.563&$\pm$&     0.039  &       0.531&$\pm$&     0.076  &       0.638&$\pm$&     0.107  \\
   1.00 ~&    30 ~  &       1.343&$\pm$&     0.019  &       1.245&$\pm$&     0.034  &       0.551&$\pm$&     0.044  &       0.443&$\pm$&     0.048  &       0.442&$\pm$&     0.095  &       0.641&$\pm$&     0.136  \\
   1.00 ~&    20 ~  &       0.903&$\pm$&     0.022  &       0.825&$\pm$&     0.043  &       0.288&$\pm$&     0.057  &       0.403&$\pm$&     0.062  &       0.413&$\pm$&     0.120  &       0.421&$\pm$&     0.125  \\
   1.00 ~&    10 ~  &       0.421&$\pm$&     0.015  &       0.337&$\pm$&     0.059  &       0.289&$\pm$&     0.077  &       0.444&$\pm$&     0.080  &       0.387&$\pm$&     0.156  &       0.444&$\pm$&     0.148  \\
   1.00 ~&     5 ~  &       0.423&$\pm$&     0.011  &       0.485&$\pm$&     0.069  &       0.390&$\pm$&     0.087  &       0.545&$\pm$&     0.089  &       0.509&$\pm$&     0.174  &       0.536&$\pm$&     0.190  \\
   1.00 ~&     0 ~  &       0.733&$\pm$&     0.010  &       0.767&$\pm$&     0.074  &       0.605&$\pm$&     0.093  &       0.652&$\pm$&     0.096  &       0.514&$\pm$&     0.190  &       0.716&$\pm$&     0.211  \\
 \hline
   2.00 ~&    40 ~  &       1.304&$\pm$&     0.019  &       1.346&$\pm$&     0.018  &       0.695&$\pm$&     0.046  &       0.514&$\pm$&     0.052  &       0.575&$\pm$&     0.076  &       ~&&     ~  \\
   2.00 ~&    30 ~  &       0.981&$\pm$&     0.021  &       0.982&$\pm$&     0.025  &       0.596&$\pm$&     0.064  &       0.499&$\pm$&     0.078  &       0.520&$\pm$&     0.095  &       ~&&     ~  \\
   2.00 ~&    20 ~  &       0.672&$\pm$&     0.021  &       0.661&$\pm$&     0.029  &       0.354&$\pm$&     0.058  &       0.369&$\pm$&     0.098  &       0.491&$\pm$&     0.121  &       ~&&     ~  \\
   2.00 ~&    10 ~  &       0.374&$\pm$&     0.016  &       0.392&$\pm$&     0.038  &       0.290&$\pm$&     0.077  &       0.366&$\pm$&     0.126  &       0.517&$\pm$&     0.157  &       ~&&     ~  \\
   2.00 ~&     5 ~  &       0.428&$\pm$&     0.013  &       0.454&$\pm$&     0.044  &       0.467&$\pm$&     0.110  &       0.461&$\pm$&     0.141  &       0.408&$\pm$&     0.174  &       ~&&     ~  \\
   2.00 ~&     0 ~  &       0.722&$\pm$&     0.013  &       0.779&$\pm$&     0.046  &       0.619&$\pm$&     0.133  &       0.709&$\pm$&     0.152  &       0.563&$\pm$&     0.192  &       ~&&     ~  \\
 \hline
\end{tabular}
 \end{center}
\caption{
\label{Table-HallC0}
Monte Carlo results for SiGe Hall Co-efficient $r_H$.  A full set
of simulation results with data for drift and Hall mobilities, 
mean energies, time of flight, diffusion, and numbers of 
blocks required can be found in \cite{Kinsler-W-2003sigeliim}
}
\end{table}

\newpage

\begin{table}
 \begin{center}
\begin{tabular}{|r|r||rcr|rcr|rcr|rcr|rcr|rcr|}
\hline
\multicolumn{20}{|c|}{  Drift Mobility $\mu_c$             } \\
\hline
 B (T)             &  Ge\%                                     & \multicolumn{18}{|c|}{$\log_{10} N_A$ (cm$^{-3}$)} \\
\hline 
          &            &       0.0 &&&       15.0 &&&       17.0 &&&       18.0 &&&       18.5 &&&       19.0 &&               \\
 \hline
   0.30 ~&    40 ~  &    1517.0&$\pm$&     3.18  &    1457.0&$\pm$&     3.26  &     931.3&$\pm$&   258.60  &     478.6&$\pm$&     0.78  &     391.3&$\pm$&     0.40  &     388.0&$\pm$&    12.14  \\
   0.30 ~&    30 ~  &    1156.0&$\pm$&     3.21  &    1116.0&$\pm$&    10.23  &     637.6&$\pm$&     1.08  &     390.4&$\pm$&     0.57  &     323.7&$\pm$&     6.41  &     305.2&$\pm$&     0.52  \\
   0.30 ~&    20 ~  &     860.7&$\pm$&     2.26  &     827.4&$\pm$&     1.94  &     498.9&$\pm$&     2.60  &     311.2&$\pm$&     1.52  &     257.4&$\pm$&     5.49  &     238.6&$\pm$&     0.30  \\
   0.30 ~&    10 ~  &     598.2&$\pm$&     1.31  &     794.7&$\pm$&   402.90  &     385.4&$\pm$&    18.17  &     240.3&$\pm$&     0.35  &     196.6&$\pm$&     0.18  &     187.9&$\pm$&    11.53  \\
   0.30 ~&     5 ~  &     516.6&$\pm$&     1.08  &     502.7&$\pm$&     1.06  &     337.1&$\pm$&     0.61  &     218.6&$\pm$&     2.44  &     177.2&$\pm$&     0.16  &     162.6&$\pm$&     0.25  \\
   0.30 ~&     0 ~  &     496.2&$\pm$&     1.20  &     481.4&$\pm$&     1.28  &     324.8&$\pm$&     0.61  &     206.4&$\pm$&     0.28  &     166.1&$\pm$&     0.16  &     149.8&$\pm$&     0.19  \\
 \hline
   1.00 ~&    40 ~  &    1309.0&$\pm$&     2.01  &    1281.0&$\pm$&     3.84  &     791.8&$\pm$&     2.10  &     477.4&$\pm$&     0.85  &     391.3&$\pm$&     1.09  &     379.0&$\pm$&     1.42  \\
   1.00 ~&    30 ~  &    1054.0&$\pm$&     1.91  &    1032.0&$\pm$&     3.09  &     633.8&$\pm$&     1.70  &     403.0&$\pm$&    25.07  &     318.3&$\pm$&     0.89  &     304.3&$\pm$&     1.16  \\
   1.00 ~&    20 ~  &     816.9&$\pm$&     1.74  &     795.1&$\pm$&     2.41  &     495.6&$\pm$&     1.36  &     309.8&$\pm$&     0.56  &     286.3&$\pm$&    64.90  &     238.2&$\pm$&     0.64  \\
   1.00 ~&    10 ~  &     587.7&$\pm$&     1.19  &     575.7&$\pm$&     5.71  &     376.0&$\pm$&     1.04  &     241.2&$\pm$&     1.92  &     196.2&$\pm$&     0.55  &     182.3&$\pm$&     0.44  \\
   1.00 ~&     5 ~  &     510.9&$\pm$&     1.04  &     507.2&$\pm$&    15.85  &     337.0&$\pm$&     0.96  &     217.3&$\pm$&     0.40  &     177.0&$\pm$&     0.50  &     162.2&$\pm$&     0.45  \\
   1.00 ~&     0 ~  &     492.2&$\pm$&     1.18  &     478.3&$\pm$&     1.46  &     324.6&$\pm$&     0.96  &     205.8&$\pm$&     0.39  &     165.6&$\pm$&     0.49  &     149.7&$\pm$&     0.43  \\
 \hline
   2.00 ~&    40 ~  &    1133.0&$\pm$&     3.04  &    1111.0&$\pm$&     3.31  &     769.2&$\pm$&     5.06  &     474.3&$\pm$&     2.23  &     389.9&$\pm$&     2.18  &       ~&&     ~  \\
   2.00 ~&    30 ~  &     946.8&$\pm$&     2.79  &     929.4&$\pm$&     3.36  &     622.3&$\pm$&     4.73  &     388.1&$\pm$&     2.21  &     317.9&$\pm$&     1.79  &       ~&&     ~  \\
   2.00 ~&    20 ~  &     761.3&$\pm$&     2.29  &     745.5&$\pm$&     2.71  &     490.7&$\pm$&     2.68  &     311.8&$\pm$&     6.37  &     251.5&$\pm$&     1.42  &       ~&&     ~  \\
   2.00 ~&    10 ~  &     567.6&$\pm$&     1.70  &     556.0&$\pm$&     2.01  &     374.7&$\pm$&     2.08  &     240.5&$\pm$&     1.37  &     195.9&$\pm$&     1.11  &       ~&&     ~  \\
   2.00 ~&     5 ~  &     499.1&$\pm$&     1.55  &     488.7&$\pm$&     1.79  &     333.9&$\pm$&     2.36  &     216.7&$\pm$&     1.26  &     178.8&$\pm$&     3.33  &       ~&&     ~  \\
   2.00 ~&     0 ~  &     482.8&$\pm$&     1.77  &     469.0&$\pm$&     1.75  &     323.6&$\pm$&     2.70  &     205.5&$\pm$&     1.24  &     165.3&$\pm$&     0.97  &       ~&&     ~  \\
 \hline
\end{tabular}
 \end{center}
 \caption{
 \label{Table-Driftm}
 Monte Carlo results for SiGe  drift mobility $\mu_c$. }
\end{table}


\newpage 

\begin{thebibliography}{99}
\bibitem[*]{e-mail}
Electronic mail: Dr.Paul.Kinsler@physics.org

\bibitem{Joelsson-FNH-1997jap}
K.B. Joelsson, Y. Fu, W.X. Ni, G.V. Hansson, \\
J. Appl. Phys. \textbf{81}, 1264 (1997), \\
\XDOI{10.1063/1.363906}.

\bibitem{Chen-LBSH-1994apl}
Y.C. Chen, S.H. Li, P.K. Bhattacharya, J. Singh, J.M. Hinckley,\\
Appl. Phys. Lett. \textbf{64}, 3110 (1994), \\
\XDOI{10.1063/1.111363}.

\bibitem{Dijkstra-1997tudelft}
J.E. Dijkstra, \\
\textit{Monte Carlo Simulation of hole transport in Si, Ge, and 
Si$_{(1-x)}$Ge$_x$}, \\
T.U. Delft (1997).

\bibitem{Dijkstra-W-1999jap}
J.E. Dijkstra, W.Th. Wenckebach, \\
J. Appl. Phys. \textbf{85}, 1587 (1999), \\
\XDOI{10.1063/1.369290}.

\bibitem{Rieger-V-1993prb}
M.M.  Rieger, P. Vogl, \\
Phys. Rev. B. \textbf{48}, 14276 (1993), \\
\XDOI{10.1103/PhysRevB.48.14276}.

\bibitem{Haye-1998tudelft}
M.J. Haye, \\
\textit{A Parallel Car-Parrinello scheme: development and applications},\\
T.U. Delft (1998).


\bibitem{Bufler-GMFK-1998jvst}
F. M. Bufler, P. Graf, B. Meinerzhagen, G. Fischer, H. Kibbel, \\
 J. of Vac. Sci. Technol. B \textbf{16}, 1667-1669 (1998); \\
ibid. \textbf{16}, 2906 (1998), \\
\XDOI{ 10.1116/1.590215}, \XDOI{10.1116/1.590293}.


\bibitem{Fischetti-L-1996jap}
M. V. Fischetti, S. E. Laux,\\
J. App. Phys. \textbf{80}, 2234-2252 (1996), \\
\XDOI{10.1063/1.363052}.
 
\bibitem{Liou-WC-1994jap}
Tsyr-Shyang Liou, Tahui Wang, Chun-Yen Chang, 
J. App. Phys. \textbf{76}, 4749-4752 (1994), 
\XDOI{10.1063/1.357244}.

\bibitem{Wenckebach-K-2002cpc}
W.Th. Wenckebach, P. Kinsler, \\
Computer Physics Communications \textbf{143}, 136 (2002), \\
XDOI{10.1016/S0010-4655(01)00440-4}.


\bibitem{MonteCarlo} 
C. Moglestu,\\
\textit{Monte Carlo simulation of semiconductor devices}, \\
(Chapman \& Hall, London, 1993).

\bibitem{Kane-1966ss}
E.O. Kane, \\
in \textit{Semiconductors and Semimetals}, 
eds. R.K. Willardson and A.C. Beer, \\
(Academic Press, New York, 1966), Vol. \textbf{1}, page 75.

\bibitem{Ridley-QPS} 
B.K. Ridley \\
\textit{Quantum Processes in Semiconductors}, \\
 (Clarendon Press, Oxford, 1988).

\bibitem{Reggiani-1985hts}
L. Reggiani, \\
in \textit{Hot electron transport in semiconductors},\\
ed. L. Reggiani, \\
(Springer-Verlag, Berlin, 1985), page 7.

\bibitem{Kinsler-W-2001jap}
P. Kinsler, W.Th. Wenckebach,\\
J. App. Phys. \textbf{90}, 1692 (2001), \\
\XDOI{10.1063/1.1384492}.

\bibitem{Kinsler-W-2000icps}
P. Kinsler, W. Th. Wenckebach, \\
Proc. 25th Internat. Conf. on the Physics of Semiconductors Parts I/II, 
ICPS-25 Osaka 2000, ed. by N. Miura, T. Ando, \\
Springer Proceedings in Physics. Vol. \textbf{87}, 
 pages 711-712 (Springer, Berlin 2001). \\
Also 
see \XARXIV{cond-mat/0201396}.

\bibitem{Dijkstra-W-1997apl}
J.E. Dijkstra, W.Th. Wenckebach, \\
Appl. Phys. Lett. \textbf{70}, 2428 (1997), \\
\XDOI{10.1063/1.118892}.

\bibitem{Briggs-WH-1998sst}
P.J. Briggs, A.B. Walker, D.C. Herbert,\\
Semicond. Sci. Tech. \textbf{13}, 680 (1998), \\
\XDOI{10.1088/0268-1242/13/7/005}.

\bibitem{Briggs-WH-1998sst2}
P.J. Briggs, A.B. Walker, D.C. Herbert,\\
Semicond. Sci. Tech. \textbf{13}, 692 (1998), \\
\XDOI{10.1088/0268-1242/13/7/006}.

\bibitem{Hinkley-S-1990}
J.M. Hinckley, J. Singh, \\
Phys. Rev. B\textbf{41}, 2912 (1990),\\ 
\XDOI{10.1103/PhysRevB.41.2912}.

\bibitem{Kinsler-W-2003sigeliim}
P. Kinsler, W.Th. Wenckebach,\\
\XDOI{10.1063/1.1622994},
\XARXIV{cond-mat/0309082}.




\end{thebibliography}
\end{document}